%% file: paper_arxiv.tex
\pdfoutput=1
\documentclass{article}

\usepackage{arxiv}
\usepackage[utf8]{inputenc} 
\usepackage[T1]{fontenc}    
\usepackage{hyperref}       
\usepackage{url}            
\usepackage{booktabs}       
\usepackage{amsfonts}       
\usepackage{nicefrac}       
\usepackage{microtype}      
\usepackage{lipsum}
\usepackage{graphicx}
\usepackage{colortbl}

\usepackage{multicol}
\usepackage{multirow}
\usepackage{booktabs}
\usepackage[table]{xcolor}
\usepackage[inline]{enumitem}
\usepackage{subcaption}
\usepackage{tabularx}
\usepackage{multirow, makecell}

\input{macro}

\title{Robust wav2vec 2.0: Analyzing Domain Shift in Self-Supervised Pre-Training}

\author{
Wei-Ning Hsu$^*$ \quad
Anuroop Sriram\thanks{Equal contribution.} \quad
Alexei Baevski \quad
Tatiana Likhomanenko \quad 
Qiantong Xu \AND
Vineel Pratap \quad
Jacob Kahn \quad
Ann Lee \quad
Ronan Collobert \quad
Gabriel Synnaeve \quad
Michael Auli
}

\begin{document}
\maketitle
\begin{abstract}
Self-supervised learning of speech representations has been a very active research area but most work is focused on a single domain such as read audio books for which there exist large quantities of labeled and unlabeled data.
In this paper, we explore more general setups where the domain of the unlabeled data for pre-training data differs from the domain of the labeled data for fine-tuning, which in turn may differ from the test data domain.
Our experiments show that using target domain data during pre-training leads to large performance improvements across a variety of setups.
On a large-scale competitive setup, we show that pre-training on unlabeled in-domain data reduces the gap between models trained on in-domain and out-of-domain labeled data by 66\%-73\%. 
This has obvious practical implications since it is much easier to obtain unlabeled target domain data than labeled data.
Moreover, we find that pre-training on multiple domains improves generalization performance on domains not seen during training.\setcounter{footnote}{0}\footnote{Code and models will be made available at \url{https://github.com/pytorch/fairseq}}
\end{abstract}

\section{Introduction}

Self-supervised learning of speech representations has received a lot attention~\cite{oord2018cpc,schneider2019wav2vec,harwath2019learning,chung2019apc,pascual2019learning} and has been demonstrated to work well both in low- and high-resource labeled data settings for automatic speech recognition (ASR)~\cite{baevski2020wav}.
However, the majority of studies focus on settings where there is little domain mismatch between the unlabeled data for pre-training, the labeled data for fine-tuning and the domain of the test data, or the \emph{target domain}.
It is well known that the performance of ASR systems trained from scratch with conventional supervised objectives can degrade significantly when tested on domains mismatched from training data~\cite{seltzer2013investigation,likhomanenko2020rethinking}.
However, the impact of domain mismatch in self-supervised speech representation learning has been much less studied.

In this paper we present a series of experiments to better understand the impact of the various domain mismatches that can occur in the self-supervised data pipeline. 
We study the effect of increasing the amount of both in-domain and out-of-domain unlabeled data when fine-tuning the resulting model on both in-domain and out-of-domain labeled data.
We also investigate the robustness of models on domains not seen during pre-training or fine-tuning.
Our findings include that adding unlabeled data whose domain matches the test data always improves performance, even if the labeled data for fine-tuning is out-of-domain. 
This has immediate practical applications since it is much easier to obtain unlabeled data for a particular target domain than labeled data.
We also find that pre-training on multiple domains increases robustness to completely unseen domains.

\section{Related Work}

This paper is related to a large body of work on robust ASR and domain adaptation.
There are two popular lines of approaches. 
The first one is feature-based, which focuses on creating robust features~\cite{stern2012features,hsu2018extracting}. Both signal processing-based~\cite{kingsbury1998robust} and learned~\cite{hsu2017disentangle} features have been explored. 
The other line is model-based, which exposes a model to diverse data while minimally pre-processing speech input in order to exploit the model capacity. 
This includes data augmentation~\cite{kim2017generation,tang2018study}, self-training on target domain~\cite{khurana2020unsupervised}, domain adversarial training~\cite{sun2017unsupervised}, and joint training~\cite{likhomanenko2020rethinking}.
The self-supervised approach explored in this paper can be categorized as model-based; however, unlike the aforementioned model-based methods, it does not require any labeled data during pre-training by using a self-supervised objective.

The most related work to this paper is~\cite{kawakami2020learning}, which investigated domain-shift for self-supervised learning, but did not dissect the domains of data used during pre-training. 
We extend this work by also examining the effect of pre-training data domain.
Furthermore, the pre-trained feature extractor in~\cite{kawakami2020learning} is fixed during supervised fine-tuning, making it more similar to feature-based approaches.
Other related work includes pre-training on multiple languages and investigating how representations transfer between languages~\cite{rivire2020unsupervised,conneau2020unsupervised}.

\section{Experimental Setup}
We experiment with wav2vec 2.0~\cite{baevski2020wav,conneau2020unsupervised} which consists of a convolutional feature encoder $f: \Inp \mapsto \Feat$ to map raw audio~$\xe$ to latent speech representations $\ze_1, \dots, \ze_T$ input to a Transformer $g: \Feat \mapsto \Context$ to output context representations $\cc_1, \dots, \cc_T$~\cite{baevski2019vqwav2vec,baevski2019effectiveness}.
Each $\ze_t$ represents about 25ms of audio strided by 20ms and the Transformer architecture follows BERT~\cite{vaswani2017transformer,devlin2018bert}.
During training, latent representations are discretized to $\zq_1, \dots, \zq_T$ with a quantization module $\Feat \mapsto \QFeat$ to represent the targets in the objective.
The quantization module uses a Gumbel softmax to choose entries from $G=2$ codebooks with $V=320$ entries each and the chosen entries are concatenated to obtain $\zq$~\cite{jegou2011ieee,jang2016gumbel,baevski2019vqwav2vec}.
The model is trained to identify the true quantized latent $\zq_t$ using $\cc_t$ for each masked time-step within a set of $K=100$ distractors $\mathbf{Q}_t$ sampled from other masked time steps.

Table~\ref{tab:exp_overview} summarizes the setups explored in each subsequent section.
Let in-domain and out-of-domain (OOD) be defined relative to the test data. 
We first study in Section~\ref{sec:add_ind_pt} whether adding in-domain data for pre-training is beneficial when OOD data is used for fine-tuning. 
In Section~\ref{sec:add_ood_pt}, we study if adding data that are OOD to pre-training can improve the performance, since previous work has only verified the effectiveness of increasing in-domain pre-training data~\cite{baevski2020wav}.
In Section~\ref{sec:unseen}, robustness of a model is evaluated through testing on domains that are not seen during pre-training or fine-tuning. 
All of the experiments mentioned above are fine-tuned on labeled data of 10 hours for low-resource setups. In Section~\ref{sec:ft_size}, a 100-hour labeled set is used for fine-tuning to validate if the conclusions from earlier sections still hold with more labeled data.

Careful ablation studies are conducted in the following two sections. 
Section~\ref{sec:pt_sim} keeps the total amount of data used in pre-training constant and varies the domain similarity by mixing OOD and in-domain data with different ratios. 
Section~\ref{sec:ind_size} dives deeper into the question raised in Section~\ref{sec:add_ind_pt} by showing how much test performance is improved with respect to the amount of in-domain pre-training data.
In the last section, we scale up the experiments by using even more data for pre-training/fine-tuning and a larger wav2vec 2.0 model to compare with previous work.

\begin{table}[h]
    \centering
    \caption{Experiment overview. Each capitalized letter denotes one domain, and ``($t$)'' is added whenever the size from that domain is of the interest for the experiments in that section.}
    \begin{tabular}{c|ccc}
        \toprule
        Sec. & Pre-Train & Fine-Tune & Test \\
        \midrule\midrule
        \ref{sec:add_ind_pt} & B vs. \{B, A\} & B/C & A \\
        \midrule
        \multirow{2}{*}{\ref{sec:add_ood_pt}}
            & B vs. \{B, C\} & A/B/C & A \\
            & A vs. \{A, C\} & A/B/C & A \\
        \midrule
        \ref{sec:unseen} & A vs. \{A, B\} & A/B/C & D/E/F \\
        \midrule
        \ref{sec:ft_size} & B vs. \{B, A\} & B($t$)/C($t$) & A \\
        \midrule
        \ref{sec:pt_sim} & B($t_1$) + A($t_2$), s.t. $t_1+t_2=const$ & A/B/C & A \\
        \midrule
        \ref{sec:ind_size} & B + A($t$) & A/B/C & A \\
        \midrule
        \ref{sec:large} & bigger \{A, B\}, bigger model & A/B & A/C \\
        \bottomrule
    \end{tabular}
    \label{tab:exp_overview}
\end{table}

\subsection{Domains and Datasets}
We consider English datasets from six domains, where datasets are regarded as from same domain if they are collected with the same process. The six domains are:
\begin{enumerate*}[label=(\arabic*)]
    \item \textbf{LibriSpeech}~\cite{panayotov2015librispeech} (\textbf{LS}) and \textbf{Libri-light}~\cite{kahn2020librilight} (\textbf{LL}), which contain about 960 and 60K hours of 16kHz crowd-sourced audiobook recordings, respectively, derived from the LibriVox project;
    \item \textbf{TED-LIUM v3}~\cite{hernandez2018ted} (\textbf{TD}), which contains 452 hours of TED conference audio in 16kHz;
    \item \textbf{Switchboard}~\cite{godfrey1993switchboard} (\textbf{SB}) and \textbf{Fisher}~\cite{cieri2005fishers,cieri2005fishert} (collectively denoted as \textbf{SF}), which consist of about 300 and 2K hours telephone conversational speech recorded in 8KHz, respectively;
    \item \textbf{Common Voice}~\cite{ardila2019common} (\textbf{CV}) with almost 700 hours crowd-sourced recordings of Wikipedia sentences in 48kHz;
    \item \textbf{Wall Street Journal} (\textbf{WS})~\cite{garofalo1993csr,wsj1} with over 80 hours of read news text recordings, and 
    \item \textbf{VoxPopuli} (\textbf{VP})~\cite{wang2021voxpopuli}, which contains 552 hours 16kHz speech recordings sourced from European Parliament plenary sessions.
\end{enumerate*}

Standard train/dev/test splits are considered for these datasets. Since SB/SF do not have a standard split, we follow the setup of~\cite{likhomanenko2020rethinking} that uses RT-03S~\cite{rt03} for validation, and Hub05 Eval2000~\cite{hub05} SwitchBoard (H-SB) and CallHome (H-CH) subsets for testing. 
The chosen datasets cover a wide variety of linguistic and acoustic domains, enabling comprehensive studies for various domain shift scenarios. 
We re-sample all datasets to 16kHz for consistency. 
Transcripts are pre-processed following~\cite{likhomanenko2020rethinking}, which upper-cases letters and removes punctuation except for apostrophes, resulting in 27+1(space)  symbols.

\subsection{Pre-training}
For experiments from Section~\ref{sec:add_ind_pt} to Section~\ref{sec:ind_size}, we consider pre-training on various combinations of LS, TD, and SF. The three largest datasets (LL, SF, CV) are combined for the scaling experiments in the last section. 
The wav2vec 2.0 \textsc{Base} architecture~\cite{baevski2019vqwav2vec} is used for all but the last section, which uses the \textsc{Large} model. \textsc{Base}/\textsc{Large} contain 12/24 transformer blocks, each of which with 768/1024 input and output dimensions, 3,072/4,096 inner (FFN) dimension and 12/16 attention heads. For \textsc{Base}, 10\% dropout is applied to the quantize/Transformer input, and output after attention, activation, and FFN layer within each transformer block. Additionally, LayerDrop~\cite{fan2019reducing} is applied with $p=5\%$. LayerNorm~\cite{ba2016layer} is used at each convolution layer in the feature encoder for both \textsc{Base} and \textsc{Large} models.

The same pre-training hyperparameters
are used for all \textsc{Base} models following~\cite{baevski2020wav} regardless what combinations of datasets they are trained on, except for the number of updates. 
Models are trained for 400K steps for the ablation studies in Section~\ref{sec:pt_sim} and \ref{sec:ind_size} for efficiency, and 800k steps elsewhere. In our preliminary studies, we found that the improvement is marginal beyond 800k.

\subsection{Fine-tuning}
We consider 10 hour subsets from LS, SF, and TD for supervised fine-tuning as the low-resource setup in most sections, 100 hour subset of TD as the mid-resource in Section~\ref{sec:ft_size}, and full LS and SB in Section~\ref{sec:large} for the high-resource setup. LS-10h is taken from the official Libri-light split,
and other subsets are sampled from the corresponding training set with genders balanced.

Pre-trained models are fine-tuned with connetionist temporal classification~\cite{graves2006connectionist}.
The same fine-tuning hyperparameters 
are used for all pre-trained models when fine-tuned on the same labeled set. These parameters are tuned using the in-domain pre-training and fine-tuning setup. For example, TD-10h parameters are selected based on the TD-dev greedy decoding word error rate (WER) of a TD pre-trained model fine-tuned on TD-10h.

\subsection{Language model and decoding}
We decode each fine-tuned model using the word-level $n$-gram language model (LM) from the target domain built with KenLM~\cite{heafield2011kenlm} and report its WER. Wav2letter++~\cite{pratap2019wav2letter} beam search decoder are used with a beam size 50, beam threshold 100, and Bayesian optimization\footnote{\url{https://github.com/facebook/Ax}} is used to find decoding hyper-parameters over 100 trials: LM weight ($[0,8]$), word score ($[-5,5]$), and silence score ($[-5,5]$).
We use the LMs provided in~\cite{likhomanenko2020rethinking} for LS, SB, TD, CV, WS, and that from~\cite{wang2021voxpopuli} for VP.

\section{Results}

\subsection{Does adding in-domain pre-training data help?}\label{sec:add_ind_pt}
Often times, it is much easier to obtain unlabeled speech for a particular domain than labeled data which requires annotation. Motivated by this, we examine the benefit of adding unlabeled in-domain data to pre-training.
To answer this question, we consider the following setup:
We pre-train models on all 7 possible combinations of LS, TD, SB, and fine-tune each of them on the ten hour subsets of the labeled version of each corpus, LS-10h, TD-10h, SB-10h, respectively. WERs on the validation sets of these three domains are presented in Table~\ref{tab:test_all}.
Unshaded columns are setups where the fine-tuning data are out-of-domain, and red numbers denotes models pre-trained with in-domain data.

For this section, we compare each black number with the red number on its right, where the red number is pre-trained additionally with the in-domain data. Both results are based on fine-tuning and testing on the same data. The benefit of adding in-domain data is clearly shown as all the red numbers are lower than the black numbers in Table~\ref{tab:test_all}.
\input{table_test_all_arxiv}

\subsection{Does adding pre-training data help if out-of-domain?}\label{sec:add_ood_pt}
Here we still pay attention to Table~\ref{tab:test_all} but compare numbers vertically. 
We split the question to be answered into two scenarios: (1) the original pre-training data does not contain in-domain data, and (2) otherwise. For the former, we compare black numbers in the second and the third row (pre-trained on one OOD dataset) with those in the last row (on two OOD ones). The black numbers in the last row are consistently better than those above, confirming the benefit of adding OOD data in this case.

For the other scenario, we compare red numbers within each column, where we found the hypothesis holds most of the time when increasing pre-training data from one domain (row 1) to two domains (row 2 and 3), with the only exception being the SB RT03 WERs when fine-tuned on SB-10h. However, when further increasing from two to three domains (row 4), the results are mixed, where about half of the cases improve.

\subsection{Does pre-training on diverse data improve robustness?}\label{sec:unseen}
We test the 21 fine-tuned models (7 pre-training dataset combinations $\times$ 3 fine-tuning datasets) from earlier sections on three domains not seen during pre-training or fine-tuning: Wall Street Journal (WS), Common Voice (CV), and VoxPopuli (VP), and report the results in Table~\ref{tab:test_ood}. In general, by comparing numbers within each column, one can observe that a model pre-trained on more domains tends to perform better than those pre-trained on fewer. To derive a summary statistic, we report the average WER over the six domains (LS, SB, TD, CV, WS, VP) for each fine-tuned model in the last three columns of Table~\ref{tab:test_ood}. It shows that pre-training on three domains achieves better performance than on two domains, which in turn is better than on one domain, regardless of what labeled data they are fine-tuned on.

\input{table_test_ood}

\subsection{Is it still effective and robust with more labeled data?}\label{sec:ft_size}
We fine-tune four models on a larger TD-100h labeled set and test on LS dev-other to verify if the conclusions from Section~\ref{sec:add_ind_pt} and \ref{sec:add_ood_pt} hold when more labeled data are available. Results in Table~\ref{tab:ft_size} confirm that both adding in-domain pre-training data (black vs. red) and adding out-of-domain data (row 1 vs row 2 vs row 3) are still effective.

\input{table_ft_100h_arxiv}

\subsection{Effect of pre-training data similarity to target domain}\label{sec:pt_sim}
Section~\ref{sec:add_ind_pt} shows that adding in-domain unlabeled data helps (red vs. black), but the improvement may be not just be due to domain similarity but it may also be due to simply increasing pre-training data size. 
To better understand the effect of domain similarity alone, we fix the amount of pre-training data to 450 hours and vary the ratio of TD/LS data to control domain similarity with respect to the test data, LS dev-other.

Table~\ref{tab:dom_sim} shows performance improvements when increasing the amount of in-domain unlabeled data up to 50\% of all pre-training data.
If there is perfect domain match for the unlabeled data, labeled data and the target domain, then more unlabeled data leads consistently to better performance (grey shaded results).
However, if the labeled data domain differs, then performance saturates at either 50\% of in-domain unlabeled data for fine-tuning with TD-10h and 75\% with SB-10h.
The effect for this is particularly strong for TD-10h and we believe that it is beneficial to have some of the unlabeled data match the domain of the labeled data for fine-tuning in order for fine-tuning to be most beneficial.
\input{table_pt_sim_arxiv}

\subsection{Effect of in-domain pre-training data size}\label{sec:ind_size}

In this section, we study the relation between the amount of in-domain data used during pre-training and performance. Two strategies are considered: 
\begin{enumerate*}[label=(\arabic*)]
    \item \textit{joint training}, which pre-trains on TD + LS of size $T$ for 400k steps, and
    \item \textit{continual training}, which first pre-trains on TD for 400k steps, and then pre-trains only on unlabeled LS of size $T$ for the numbers of steps specified in \autoref{tab:ind_size}. 
\end{enumerate*}
In practice, it is convenient to pre-train a model on a large dataset, and then adapt it to a new domain of interest by running additional pre-training steps on that domain. 

Result in \autoref{tab:ind_size} show that adding more in-domain unlabeled data continually improves performance for both joint training and continual training, and both strategies achieve similar performance.
Compared to \autoref{tab:dom_sim}, using all in-domain unlabeled data still performs well when fine-tuned on TD-10h, because pre-training always includes all TD data which matches the domain of labeled data for fine-tuning.
\input{table_indom_size_arxiv}

\input{table_large_models_arxiv}

\subsection{Larger model, more pre-training and fine-tuning data}\label{sec:large}

Finally, we pre-train a single large wav2vec 2.0 model with 300M parameters~\cite{baevski2020wav} on three domains (LL, SF and CV) for 800K steps and fine-tune it on 10 hours as well as all data of the LS and SB datasets. We evaluate each model on the validation and test splits for LS, SB, CV and TD, showing in-domain and OOD performance. We compare our model to supervised models trained on single or multiple domains reported in~\cite{likhomanenko2020rethinking} (Table~\ref{tab:large_models}). 

On in-domain data (LS/SB), our model achieves superior performance to all single- and multi-domain models in~\cite{likhomanenko2020rethinking} except on the CallHome (H-CH) set.
Pre-training on multiple domains is particularly effective when testing on domains different from fine-tuning: when fine-tuned on LS, WER is reduced by a relative 35\% to 50\% on SB, CV, TD compared to the single-dataset baseline trained on full LS in~\cite{likhomanenko2020rethinking}.
Moreover, we even achieve better OOD performance compared to that baseline when fine-tuning on just 10 hours of LS data (LS-10h). The same trend holds when comparing fine-tuning on SB-10h and for the baseline in~\cite{likhomanenko2020rethinking} trained on all of SF (200x more labeled data).
Our model is not pre-trained or fine-tuned on any TD data, and when fine-tuning it with LS, it achieves better performance on TD than the supervised SOTA trained on TD~\cite{zhou2020rwth} as well as the joint RASR model trained on five labeled datasets including TD~\cite{likhomanenko2020rethinking}.

\subsection{Effectiveness of pre-training}

In the supervised learning paradigm, practitioners who would like to build a system for a new domain, can either train on existing OOD labeled data or build a corpus of labeled data in the new domain.
With pre-training, we have a third option: collect unlabeled data in the new domain and fine-tune on existing labeled OOD data.
This has the clear advantage of unlabeled in-domain being often much easier to obtain than transcribed in-domain data.

To get a better sense of how effective this third choice is, we re-examine our earlier experiments (\autoref{tab:test_all}). 
We measure the performance gap between the ideal setting where we have access to the labels of the in-domain data (Topline; PT-on-all, FT-on-InD) and a setting where we have only access to labeled out-of-domain data (Baseline; (PT,FT)-on-OOD).
\autoref{tab:remark} shows how much of this gap is closed by the third option which in this case is pre-training on a variety of domains, including in-domain unlabeled data, while fine-tuning on labeled OOD data (Proposed; PT-on-all, FT-on-OOD).
Pre-training on in-domain data closes at least 73\% of the gap across a variety of settings.

We repeat the same analysis for the much larger and more competitive systems presented in \autoref{sec:large}.
Performance is measured on the evaluation sets of Switchboard (RT03/H-SB/H-CH).
As baseline, we consider the competitive supervised model of~\cite{likhomanenko2020rethinking} trained on labeled LS only (Baseline; No PT, FT-on-OOD).
\autoref{tab:remark2} shows that pre-training on unlabeled in-domain data and fine-tuning on OOD closes between 66\%-73\% of the performance gap between the Topline and the Baseline performance.
This bodes very well for practitioners who would like to build a model for a new domain since it is generally much easier to obtain unlabeled data for a new domain compared to transcribed data.

\input{table_final_remark}

\section{Conclusion}

We present the first controlled study to better understand domain-shift in self-supervised learning for ASR. 
Results show that adding unlabeled in-domain data improves performance, even when the fine-tuning data does not match the test domain.
With no access to in-domain labeled data, pre-training on unlabeled in-domain data closes 66-73\% of the performance gap between the ideal setting of in-domain labeled data and a competitive supervised out-of-domain model.
Moreover, self-supervised representations trained on a variety of domains are robust and lead to better generalization performance on domains completely unseen during pre-training and fine-tuning.
Retaining some unlabeled data from the same domain as the fine-tuning data is beneficial though.

\clearpage
\bibliographystyle{unsrt}  
\bibliography{refs}

\end{document}

%% file: macro.tex
\newcommand{\Inp}{\mathcal{X}}
\newcommand{\Feat}{\mathcal{Z}}
\newcommand{\QFeat}{\mathcal{Q}}
\newcommand{\Context}{\mathcal{C}}

\newcommand{\xe}{\mathbf{x}}
\newcommand{\ze}{\mathbf{z}}
\newcommand{\zq}{\mathbf{q}}

\newcommand{\cc}{\mathbf{c}}

%% file: table_test_all_arxiv.tex
\begin{table}[ht]
    \centering
    \caption{Validation WER on TD, LS, and SB of models pre-trained (PT) on various subsets of \{TD, LS, SB\}, and fine-tuned (FT) on TD-10h, LS-10h, or SB-10h.}\label{tab:test_all}
    
    \begin{tabular}{c|>{\columncolor[gray]{0.9}}c>{\columncolor[gray]{0.9}}c|cc|cc}
        \toprule
        \multirow{3}{*}{X}
        & \multicolumn{6}{c}{\textit{TED-LIUM (TD) dev WER}} \\
        & \multicolumn{2}{>{\columncolor[gray]{0.9}}c|}{FT on TD-10h}
        & \multicolumn{2}{c|}{FT on LS-10h} 
        & \multicolumn{2}{c}{FT on SB-10h} \\
        & PT on X & X+TD & PT on X & X+TD & PT on X & X+TD \\
        \midrule
        None  &     diverge & {\color{red}9.93} & diverge     & {\color{red}10.99} & diverge     & {\color{red}11.32} \\
        SF    & 12.12 & {\color{red}9.60} & 14.82 & {\color{red}11.08} & 99.63 & {\color{red}11.04} \\
        LS    &  9.81 & {\color{red}8.59} & 12.92 & {\color{red} 8.91} & 13.08 & {\color{red}10.39} \\
        SF+LS &  9.13 & {\color{red}8.91} & 10.61 & {\color{red} 9.67} & 12.25 & {\color{red}10.75} \\
        \bottomrule
    \end{tabular}
    \vspace{.5em}
    
    \begin{tabular}{c|cc|>{\columncolor[gray]{0.9}}c>{\columncolor[gray]{0.9}}c|cc}
        \toprule
        \multirow{3}{*}{X}
        & \multicolumn{6}{c}{\textit{LibriSpeech (LS) dev-other WER}} \\
        & \multicolumn{2}{c|}{FT on TD-10h} 
        & \multicolumn{2}{>{\columncolor[gray]{0.9}}c|}{FT on LS-10h}
        & \multicolumn{2}{c}{FT on SB-10h} \\
        & PT on X & X+LS & PT on X & X+LS & PT on X & X+LS \\
        \midrule
        None  & diverge     & {\color{red}14.60} & diverge     & {\color{red}10.53} & diverge     & {\color{red}17.92} \\
        SF    & 28.91 & {\color{red}14.30} & 20.36 & {\color{red}10.44} & 94.38 & {\color{red}15.53} \\
        TD    & 23.44 & {\color{red}12.81} & 15.36 & {\color{red} 9.71} & 27.50 & {\color{red}15.46} \\
        SF+TD & 20.50 & {\color{red}13.58} & 14.42 & {\color{red}10.39} & 21.99 & {\color{red}13.89} \\
        \bottomrule
    \end{tabular}
    \vspace{.5em}
    
    \begin{tabular}{c|cc|cc|>{\columncolor[gray]{0.9}}c>{\columncolor[gray]{0.9}}c}
        \toprule
        \multirow{3}{*}{X}
        & \multicolumn{6}{c}{\textit{Switchboard (SB) RT03 WER}} \\
        & \multicolumn{2}{c|}{FT on TD-10h} 
        & \multicolumn{2}{c|}{FT on LS-10h} 
        & \multicolumn{2}{>{\columncolor[gray]{0.9}}c}{FT on SB-10h} \\
        & PT on X & X+SF & PT on X & X+SF & PT on X & X+SF \\
        \midrule
        None  & diverge     & {\color{red}18.90} & diverge     & {\color{red}19.30} & diverge     & {\color{red}10.80} \\
        TD    & 35.70 & {\color{red}16.20} & 34.60 & {\color{red}17.40} & 18.70 & {\color{red}11.00} \\
        LS    & 33.60 & {\color{red}17.80} & 36.50 & {\color{red}16.10} & 18.20 & {\color{red}11.00} \\
        TD+LS & 29.70 & {\color{red}17.40} & 28.90 & {\color{red}16.90} & 15.60 & {\color{red}10.80} \\
        \bottomrule
    \end{tabular}
    \vspace{-.5em}
\end{table}

%% file: table_test_ood.tex
\begin{table*}[t]
    \centering
    \caption{Validation WER on domains unseen during pre-training or fine-tuning: WS, CV, and VP. \{TD, LS, SB\}-10h denote the labeled data used for fine-tuning for the corresponding column. We also report the average WER over the dev sets of WS, CV, VP, TD, LS, SB.}
    \label{tab:test_ood}
    \resizebox{.9\linewidth}{!}{
    \begin{tabular}{c|ccc|ccc|ccc|ccc}
        \toprule
        \multirow{2}{*}{PT on X} 
        & \multicolumn{3}{c|}{\textit{WS-dev WER}}
        & \multicolumn{3}{c|}{\textit{CV-dev WER}}
        & \multicolumn{3}{c|}{\textit{VP-dev WER}}
        & \multicolumn{3}{c}{\textit{Avg WER (over 6 devs)}}\\
        & TD-10h & LS-10h & SB-10h & TD-10h & LS-10h & SB-10h
        & TD-10h & LS-10h & SB-10h & TD-10h & LS-10h & SB-10h \\
        \midrule
        TD       & 11.32 &  9.87 & 11.10 & 36.93 & 34.53 & 46.04 & 18.59 & 16.67 & 22.41 & 22.65 & 20.34 & 22.85 \\
        LS       &  9.62 &  9.18 & 10.10 & 31.80 & 31.49 & 43.63 & 20.47 & 18.50 & 27.64 & 19.98 & 19.85 & 21.76 \\
        SF       & 14.65 & 12.79 & 99.25 & 44.14 & 43.08 & 94.53 & 22.88 & 24.73 & 99.97 & 23.60 & 22.51 & 83.09 \\
        \midrule
        TD+LS    &  9.08 &  8.00 &  8.95 & 28.54 & 27.34 & 37.59 & 14.77 & 14.43 & 17.77 & 17.25 & 16.21 & 17.63 \\
        TD+SF    & 10.64 &  9.83 & 10.01 & 32.98 & 32.02 & 36.09 & 16.19 & 16.69 & 17.25 & 17.69 & 16.90 & 17.90 \\
        LS+SF    &  9.76 &  8.72 &  9.32 & 28.67 & 28.29 & 34.12 & 15.49 & 16.18 & 19.60 & 15.86 & 15.06 & 16.97 \\
        \midrule
        TD+LS+SF &  9.13 &  8.44 &  8.94 & 28.44 & 27.13 & 30.92 & 15.03 & 15.44 & 16.91 & 15.42 & 14.66 & 15.37 \\
        \bottomrule
    \end{tabular}
    }
    \vspace{-1em}
\end{table*}

%% file: table_ft_100h_arxiv.tex
\begin{table}[h]
    \centering
    \caption{Effect of more labeled data (LS dev-other WER).}
    \label{tab:ft_size}
    \begin{tabular}{c|cc|cc}
        \toprule
        \multirow{3}{*}{X}
        & \multicolumn{4}{c}{\textit{LibriSpeech (LS) dev-other WER}} \\
        & \multicolumn{2}{c|}{FT on TD-100h} 
        & \multicolumn{2}{c}{FT on TD-10h} \\
        & PT on X & X+LS & PT on X & X+LS \\
        \midrule
        SF    & 19.45 & {\color{red}10.98} & 28.91 & {\color{red}14.30} \\
        TD    & 18.35 & {\color{red}9.84}  & 23.44 & {\color{red}12.81} \\
        SF+TD & 15.30 & {\color{red}10.57} & 20.50 & {\color{red}13.58} \\
        \bottomrule
    \end{tabular}
\end{table}

%% file: table_pt_sim_arxiv.tex
\begin{table}[h]
    \centering
    \caption{Controlling for the amount of pre-training data. We show Librispeech dev-other validation WER with exactly 450h of pre-training data.}\label{tab:dom_sim}
    \begin{tabular}{cc|c>{\columncolor[gray]{0.9}}cc}
        \toprule
        \multicolumn{2}{c|}{Pre-train Size} 
        & \multicolumn{3}{c}{\textit{LS dev-other WER}, FT on} \\
        TD & LS & TD-10h & LS-10h & SB-10h \\
        \midrule
        450.0h & None   & 23.06 & 16.08 & 28.72 \\
        337.5h & 112.5h & 15.97 & 12.61 & 20.83 \\
        225.0h & 225.0h & 14.46 & 11.52 & 18.27 \\
        112.5h & 337.5h & 14.15 & 11.00 & 18.83 \\
        None   & 450.0h & 15.74 & 10.75 & 18.86 \\
        \bottomrule
    \end{tabular}
\end{table}

%% file: table_indom_size_arxiv.tex
\begin{table}[h]
    \centering
    \caption{Effect of in-domain pre-training size in terms of LibriSpeech dev-other WER. Joint models were pre-trained on TD + LS of size $T$ for 400k steps. Continual models were first pre-trained only on TD, and then pre-trained only on unlabeled LS of size $T$. Numbers of continual pre-traing steps are included.}
    \label{tab:ind_size}
    \begin{tabular}{cc|c|cc|cc}
        \toprule
        \multicolumn{2}{c|}{Pre-train Size} & Num. of & \multicolumn{4}{c}{\textit{LS dev-other WER}} \\
        \multirow{2}{*}{TD} & \multirow{2}{*}{LS} & Cont. PT Steps &  \multicolumn{2}{c|}{FT on TD-10h} & \multicolumn{2}{c}{FT on SB-10h} \\
        & & on LS & Joint & Cont. & Joint & Cont. \\
        \midrule
        \multirow{6}{*}{450h}
        & None   & -    & \multicolumn{2}{c|}{23.06} & \multicolumn{2}{c}{28.72} \\
        & 10m    & 10k  & 22.93 & 22.75 & 29.24 & 29.35 \\
        & 1h     & 25k  & 22.09 & 22.25 & 28.89 & 28.77 \\
        & 10h    & 50k  & 21.83 & 21.77 & 26.78 & 27.59 \\
        & 100h   & 100k & 18.80 & 18.82 & 23.55 & 24.18 \\
        & 960h   & 400k & 13.19 & 13.91 & 16.75 & 17.03 \\
        \bottomrule
    \end{tabular}
\end{table}

%% file: table_large_models_arxiv.tex
\begin{table*}[h]
    \centering
    \caption{Validation and test WER of \textsc{Large} models pre-trained on LL+SF+CV and fine-tuned on subsets of LS and SB. For comparison, $n$-gram decoding results from \cite{likhomanenko2020rethinking} are included (note test sets are donwsampled to 8kHz in~\cite{likhomanenko2020rethinking}). SOTA results with supervised training on in-domain datasets are reported in the bottom rows (these models are not tested on OOD data).}
    \label{tab:large_models}
    \resizebox{\linewidth}{!}{    \begin{tabular}{cc|cccc|ccc|cc|cc}
        \toprule
        \multirow{2}{*}{PT} & \multirow{2}{*}{FT} &
        \multicolumn{4}{c|}{LibriSpeech} 
        & \multicolumn{3}{c|}{Switchboard} 
        & \multicolumn{2}{c|}{CommonVoice} 
        & \multicolumn{2}{c}{TED-LIUM} \\
        & & dev-c & dev-o & test-c & test-o 
        & RT03 & H-SB & H-CH 
        & valid & test & valid & test \\
        \midrule
        \multirow{4}{*}{LL+SF+CV}
        & LS-10h & 2.78	& 5.78 & 3.17 & 6.26 & 17.1 & 13.1 & 19.2 &17.30 &21.06 & 7.11 & 7.65 \\
        & LS     & 1.77 & 3.84 & 2.08 & 4.15 & 13.0 &  9.8 & 14.6 &12.36 &15.27 & 5.06 & 5.38 \\
        & SB-10h & 3.83	& 7.76 & 4.12 & 7.96 &  9.8 &  6.3 & 13.3 &20.32 &24.69 & 7.01 & 6.82 \\
        & SB     & 3.03 & 6.74 & 3.30 & 7.08 &  7.7 &  4.9 &  9.9 &18.02 &22.48 & 5.62 & 5.18 \\
        \midrule
        \multirow{3}{*}{None~\cite{likhomanenko2020rethinking}}
        & LS    & 2.0 &  5.3 & 2.5 &  5.6 & 27.5 & 19.3 & 26.4 & 18.8 & 22.5 & 7.8 & 9.4 \\
        & SF    & 7.1 & 19.1 & 7.9 & 20.4 & 10.4 &  6.5 & 10.3 & 31.7 & 36.0 & 8.5 & 8.8 \\
        & Joint (LS+SF+CV+TD+WS)
        & 2.0 &  5.2 & 2.5 &  5.6 &  9.8 &  5.9 &  9.5 & 10.5 & 12.6 & 5.4 & 5.7 \\
        
        \midrule
        \multicolumn{13}{c}{\textit{State-of-the-art Single-Domain Supervised Models}} \\
        None~\cite{gulati2020conformer}        & LS   & - & - & \textit{1.9} & \textit{3.9} & - & - & - & - & - & - & - \\
        None~\cite{wang2020investigation}      & SB & - & - & - & - & \textit{11.4} & \textit{6.3} & \textit{13.3} & - & - & - \\
        None~\cite{likhomanenko2020rethinking} & CV & - & - & - & - & - & - & - & \textit{10.7} & \textit{13.6} & - & - \\
        None~\cite{zhou2020rwth}               & TD & - & - & - & - & - & - & - & - & - & \textit{5.1} & \textit{5.6} \\
        \bottomrule
    \end{tabular}
    }
\end{table*}

%% file: table_final_remark.tex
\begin{table}[ht]
    \centering
    \caption{Word error rate reduction (WERR) derived from Table~\ref{tab:test_all} to assess the effectiveness of pre-training on in-domain unlabeled data and fine-tuning on OOD labeled data (Proposed): we measure how much of the performance gap Proposed closes with respect settings where only OOD data is available (Baseline) and the ideal setting where the labels for the in-domain data is available (Topline). WERR is defined as $(proposed - baseline) / (topline - baseline)$.}\label{tab:remark}
    
    \resizebox{\linewidth}{!}{
    \begin{tabular}{ccc|ccc|c}
        \toprule
        \multirow{2}{*}{Test On} & \multicolumn{2}{c|}{Fine-tune data} 
        & Topline & Proposed & Baseline & \multirow{2}{*}{WERR} \\
        & InD & OOD & PT-on-all, FT-on-InD & PT-on-all, FT-on-OOD & (PT,FT)-on-OOD \\
        \midrule
        \multirow{2}{*}{TD dev} & \multirow{2}{*}{TD-10h}
          & LS-10h & \multirow{2}{*}{8.91} & 9.67  & 12.92 & 81.1\%\\
        & & SB-10h &                       & 10.75 & 99.63 & 98.0\%\\
        \midrule
        \multirow{2}{*}{LS dev-other} & \multirow{2}{*}{LS-10h}
          & TD-10h & \multirow{2}{*}{10.39} & 13.58 & 23.44 & 75.6\%\\
        & & SB-10h &                        & 13.89 & 94.38 & 95.8\%\\
        \midrule
        \multirow{2}{*}{SB RT03} & \multirow{2}{*}{SB-10h}
          & TD-10h & \multirow{2}{*}{10.80} & 17.40 & 35.70 & 73.5\%\\
        & & LS-10h &                        & 16.90 & 36.50 & 76.3\%\\
        \bottomrule
    \end{tabular}
    }
\end{table}

\begin{table}[ht]
    \centering
    \caption{Word error rate reduction (WERR) derived from Table~\ref{tab:large_models} for the high-resource setup and large models. WERR is defined as $(proposed - baseline) / (topline - baseline)$ (cf.~\autoref{tab:remark}).}\label{tab:remark2}
    
        \begin{tabular}{ccc|ccc|c}
        \toprule
        \multirow{2}{*}{Test On} & \multicolumn{2}{c|}{Fine-tune data} 
        & Topline & Proposed & Baseline & \multirow{2}{*}{WERR} \\
        & InD & OOD & PT-on-all, FT-on-InD & PT-on-all, FT-on-OOD & No PT, FT-on-OOD~\cite{likhomanenko2020rethinking} \\
        \midrule
        SB RT03 & SB & LS & 7.7 & 13.0 & 27.5 & 73.2\% \\
        SB H-SB & SB & LS & 4.9 &  9.8 & 19.3 & 66.0\% \\
        SB H-CH & SB & LS & 9.9 & 14.6 & 26.4 & 71.5\% \\
        \bottomrule
    \end{tabular}
\end{table}